\documentclass[twocolumn,english,prx,superscriptaddress]{revtex4-1}
\usepackage[utf8]{inputenc}
\usepackage{color}
\usepackage{babel}
\usepackage{bm}
\usepackage{graphicx}
\usepackage{physics}
\usepackage{amsmath}
\usepackage{amssymb}
\usepackage{empheq}
\usepackage{times}
\usepackage{hyperref}
\usepackage{comment}

\begin{document}

\title{Speed limits of two-qutrit gates}

\author{Bora Basyildiz}
\email{bbasyildiz@caltech.edu}
\affiliation{Institute for Quantum Information and Matter and Department of Physics, California Institute of Technology, Pasadena, CA 91125, USA}

\author{Zhexuan Gong}
\affiliation{Department of Physics, Colorado School of Mines, Golden, Colorado 80401, USA}

\author{Sahel Ashhab}
\email{ashhab@nict.go.jp}
\affiliation{Advanced ICT Research Institute, National Institute of Information and Communications Technology,
4-2-1, Nukui-Kitamachi, Koganei, Tokyo 184-8795, Japan}
\affiliation{Research Institute for Science and Technology, Tokyo University of Science, 1-3 Kagurazaka,
Shinjuku-ku, Tokyo 162-8601, Japan}

\begin{abstract}
      
The speed of elementary quantum gates sets a limit on the speed at which quantum circuits can be applied and, as a result, the size of the computations that can be performed on a quantum computer \cite{Howard2023,Basyildiz2023,Ashhab2022,Basyildiz2024Optimal}. This limitation stems from the fact that present-day quantum hardware systems have finite coherence times that limit the total computation time \cite{Bluvstein2025,Moses2023,Teoh2023}. The speeds of qubit gates in various hardware settings have been well studied over the past few decades. The recent interest in multi-level quantum systems naturally creates a need for similar investigations of the speeds of multi-level or qudit gates \cite{Campbell_2014,Sur-Kolay_2018,Muschik_2025,Basyildiz2024Optimal,Ashhab2012}. In this work, we perform an empirical study of the speed limit for the three-level or qutrit CZ gate. Our analysis focuses on a theoretical model for capacitively coupled superconducting transmons but can be extended to other systems. We generate CZ gate protocols using optimal control theory techniques and observe when the fidelity crosses certain thresholds. In addition to the empirical approach, we derive an analytical speed limit for the qutrit CZ gate using traditional quantum speed limit techniques. We compare the speed limits derived using these two different approaches and discuss the gap that remains between them. We also compare the time needed to implement the qutrit CZ gate with its qubit counterpart.
\end{abstract}

\maketitle

\section{Introduction}
    As the field of quantum computation progresses and transitions from the noisy intermediate scale quantum (NISQ) computer era \cite{Preskill2018} to one where millions of quantum operations can be implemented (Megaquop) \cite{Preskill2025}, the ability of quantum computers to run fast high-fidelity gates has become a critical issue for academia and industry alike. Demonstrating a computational advantage for most quantum algorithms, such as through Shor's Algorithm, Grover's Algorithm, Quantum Phase Estimation, Quantum Chemistry Physics Simulations, and others\cite{Shor_1997,Kitaev_1995,Aspuru-Guzik_2014,Yelin_2025,Regev_2024,Grover_1996,Thompson_2021,Aspuru-Guzik_2020,Chan_2020,Chan_2022}, requires significantly higher circuit depths than what is currently possible. Faster individual gates will enable us to achieve higher circuit depths for finite-coherence hardware.
    
    Leveraging higher energy states has been recently used for a wide variety of quantum applications, including magic state distillation \cite{Campbell_2014,Browne_2012}, quantum error correction \cite{Sur-Kolay_2018,Muralidharan_2017,Zeng_2018}, and the quantum simulation of $d$-level systems \cite{Muschik_2025,Yelin_2025,Aspuru-Gusik_2019,Tan_2024,Haldane_1983,Guerreschi_2020}. In an analogy to binary two-level quantum systems being called qubits, multi-level systems with $d$ quantum states are called qudits. Three-level systems are specifically called qutrits. To take advantage of qutrits in quantum computation, one must implement a universal set of qutrit operations with high fidelity \cite{Yash2025,McCord2025}. An important qutrit operator that can serve as the entangling gate for a universal gate set is the qutrit CZ gate. This gate was recently demonstrated with high fidelity using superconducting transmon systems \cite{Goss2022,Luo2023}. An important question in this context is whether the gate can be performed faster than the speeds achieved in these recent experiments. The theoretical limits on the achievable speeds for qutrit gates have not been adequately studied to the best of our knowledge.
    
    In contrast to qutrits, the speed limits of two-qubit gates are well understood \cite{Vidal2002,Khaneja2001,Dur2001}. An accurate speed limit, the fundamental minimum amount of time necessary to generate the gate with high fidelity, can be derived for most standard two-qubit gates. These speed limits are typically determined by the coupling strength between the qubits, since local operations for the individual qubits can be performed much faster than two-qubit gates and are often treated as instantaneously fast \cite{Howard2023,Basyildiz2024Optimal}.  Furthermore, the speed limits of two-qubit gates in higher-level systems have been studied, and it has been shown that one can leverage the additional higher levels to achieve faster two-qubit gates \cite{Basyildiz2023,Basyildiz2024Optimal,Ashhab2022}. 
    
    In this paper, we provide a few new results pertaining to the speed limits for two-qutrit gates. We focus on the qutrit CZ gate in a system of two capacitively coupled superconducting transmon qutrits,3which will allow us to compare our results with recent experimental results. Furthermore, superconducting transmons are among the leading quantum platforms used for quantum computing hardware \cite{Google2025,Putterman2025,Moreno2025,Teoh2023,Li2024,Rosen2025}. However, one of the main challenges that limit their performance is their finite coherence time, making it necessary to minimize gate times \cite{Ganjam2024,Burnett2019}. In addition, the static capacitive coupling architecture is common in superconducting systems, and current experimental demonstrations of qutrit CZ gates leverage this coupling scheme. Our theoretical analysis will therefore focus on this architecture and coupling. 
    
    To study the speed limits of the qutrit CZ gate, we will numerically generate fidelity-optimized CZ gates for a given time $T$ through optimal control theory. Then we will observe the point at which the fidelity $F$ of our gates crosses a preset high-fidelity threshold such as $F \geq 99.9\%$. This point represents an empirically observed speed limit. It defines the minimum time at which we know with certainty that high-fidelity gates can be generated.
    
    In addition to our optimal control protocols, we will use traditional quantum speed limit (QST) methodologies to derive a theoretical speed limit for the qutrit CZ gate. Two of the methods often used to bound quantum speed limits are based on the Mandelstam-Tamm and Margolus-Levitin inequalities, which bound the time for an orthogonal state transfer \cite{Frey2016,Mandelstam1991,Margolus1998}. To directly bound the time to generate two-qubit unitaries, one uses the canonical decomposition given by Ref. \cite{Vidal2002}. An extension of these bounds to multi-level systems can be used to bound the time for two-qudit gates to be applied in multi-level systems \cite{Basyildiz2023,Basyildiz2024Optimal}. We find a discrepancy between the theoretical and empirically observed speed limits, which suggests that there could be a tighter theoretical bound compared to the one that we present here. 
    
    The rest of the paper is organized as follows: We will first generate optimal control protocols to determine an approximate speed limit for a qutrit CZ gate by observing the time in which a high-fidelity gate can be generated. Then we will use quantum speed limit techniques to derive a theoretical bound on the speed limit, and discuss the gap between the speed of the optimal protocol and the theoretical speed limit.

\section{Optimal Control}
    Optimal Control Theory (OCT) techniques optimize control pulses to achieve various objectives in quantum systems. Example objectives are fidelity, time, and leakage \cite{Basyildiz2023,Ashhab2022,Trowbridge2023,Abdelhafez2020,Glaser2015,Halaski2024,Koch2022,Koch2021,Larrouy2020,Fischer2019,Dalgaard2022,Vetter2024,Gambetta2011,Wittler2021,Werninghaus2021,Machnes2018,Peng2023}. Areas in which OCT has been applied include two-qubit and multi-level system control, spin dynamic control, and nuclear magnetic resonance \cite{Glaser2015,Halaski2024,Werschnik2007,Sporl2007,Muller2011,Reich2012,Huang2014}. In superconducting qutrit experiments, the pulses are often a combination of microwave signals that drive the various transitions in the system. While the set of pulses generated by an OCT calculation are not guaranteed, or in fact likely, to converge to the global optimum, in many of the applications that we are concerned with, the pulses converge to a sufficiently good local minima when given an ample number of iterations and random seed initializations \cite{Chambers1965,Lee1967,bryson1975}. Hence OCT has historically proven to be an efficient way to solve optimization problems.

\begin{figure}[ht!]
       \centering
       \includegraphics[width=0.5\textwidth]{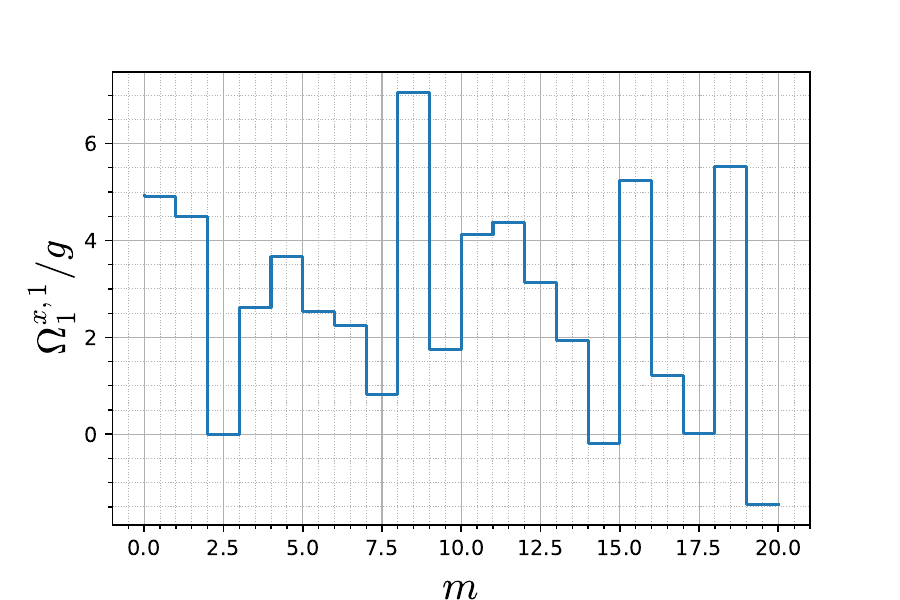}
       \caption{An example of the optimized pulse $\Omega_1^{x,1}$ representing the $x$ drive for the control pulse that is resonant with the $\ket{0}\leftrightarrow\ket{1}$ transition for the first qutrit. We plot the first 20 out of 40 segments of our protocol for the point $T/T_{min} = 1.1$ for the qutrit CZ gate (see Fig. \ref{fig:qutritCZ_plot}).}
       \label{fig:pulses}
   \end{figure}

    Our OCT protocol is derived from the open-source project QuOpt developed in \cite{Basyildiz2023,Howard2023,BasyildizGithub2023,Basyildiz2024Optimal}. The cost function used for optimization is the gate infidelity $r = 1 -F$ between the target gate. In our system, this will be the qutrit CZ gate, and we will calculate the gate fidelity of the unitary generated from our QuOpt protocol to the qutrit CZ gate. To determine the infidelity $r$, we generate the average gate fidelity $F$ for a given gate by taking its fidelity over the generators of our Hilbert space, such that \cite{Nielsen2002,Howard2023,Basyildiz2023,Basyildiz2024Optimal}

    \begin{equation}\label{eq: fidelity}
        F = \frac{\sum_j \mathrm{tr} \left(\hat{U} \hat{U}_j^\dagger \hat{U}^\dagger \hat{\mathcal{U}}\hat{U}_j \hat{\mathcal{U}}^\dagger\right) + d^2}{d^2(d+1)}
    \end{equation}
    
    \noindent where $\hat{U}_j$ represent the generators for our Hilbert space with dimension $d$. The matrix $\hat{U}$ is the target gate (in our case qutrit CZ) and $\hat{\mathcal{U}}$ is the gate generated from our QuOpt protocol. For our two-qutrit space, we will set $d = 9$ with the generators being of the form $\hat{\lambda_j}\otimes\hat{\lambda_k}$ where $\hat{\lambda_j},\hat{\lambda_k}$ are the single-site Gell-Mann matrices for the two qutrits. Now, to generate our QuOpt protocol unitary $\hat{\mathcal{U}}$, we have the formula

    \begin{equation}\label{eq: unitary evolution}
        \hat{\mathcal{U}} = \mathcal{T}e^{-i\int_0^T\hat{H}(t)dt}
    \end{equation}

    \noindent where $\mathcal{T}$ is the time-ordering operator for the evolution from time $t=0$ to the total time $t=T$. We partition the Hamiltonian into two terms: one that describes the static capacitive coupling and one that describes the time-dependent control pulses such that $ \hat{H} = \hat{H}_0 + \hat{H}_1(t)$. The approximate coupling Hamiltonian in the absence of external fields is given by \cite{Ashhab2022,Howard2023,Basyildiz2024Optimal}

    \begin{equation}\label{eq:H0}
        H_0 = g(\hat{a}_1 + \hat{a}_1^\dagger)\otimes (\hat{a}_2 + \hat{a}_2^\dagger),
    \end{equation}

   \noindent where $g$ is the coupling strength. Our time-dependent control Hamiltonian is described by 

   \begin{equation}\label{eq:control_hamiltonian}
       \hat{H}_1(t) = \sum_{i=1,2} \sum_{j=0}^{d-2} \sum_{\gamma=x,y} \Omega_i^{\gamma,j} (t) \hat{\sigma}_i^{\gamma,j},
   \end{equation}
    
   \noindent where the index $i$ labels the two qudits, $j$ labels the quantum states of each qudit, and $\gamma$ represents the Pauli $x$ and $y$ (or, in other words, sine and cosine) drives for the transition $\ket{j}\leftrightarrow\ket{j+1}$. For example, $\hat{\sigma}_1^{x,0}$ has matrix elements equal to 1 for the transition $\ket{0}\leftrightarrow\ket{1}$ in qudit 1 and zero otherwise. This operator can be thought of as X quadrature drive for the $\ket{0}\leftrightarrow\ket{1}$ transition. Note that our drives will only be between neighboring states. As mentioned above, in the case of qutrits, the operators $\hat{\sigma}_i^{\gamma,j}$ are represented by the Gell-Mann matrices for the respective transitions. It is also worth noting that Eq.~\ref{eq:control_hamiltonian} has independent drives for the different transitions, e.g.~$\ket{0}\leftrightarrow\ket{1}$ and $\ket{1}\leftrightarrow\ket{2}$ in the qutrit case. We optimize over the Rabi frequencies $\Omega_i^{\gamma,j} (t)$ of the microwave fields that are resonant with the respective driven transitions in our QuOpt protocol. We implement the drives in a segmented fashion such that our total time is discretized into $M$ total segments where each segment has time $t\in [\frac{m-1}{M}T,\frac{m}{M}T]$. Now, our drives $\Omega_i^{\gamma,j}(t)$ will be static over a given segment $m$. Thus, the unitary operator in Eq.~\ref{eq: unitary evolution} can be expressed in the simpler form 

   \begin{equation}
       \hat{U} = \hat{U}_1 \hat{U}_2\ldots \hat{U}_M,
   \end{equation}

   \noindent where, for a given segment $m$ with unitary evolution $U_m$, we have 

   \begin{equation}
       \hat{U}_m = e^{-i(\hat{H}_0 + \hat{H}_{1,m})\frac{T}{M}}.
   \end{equation}

   \noindent where the drive term of the Hamiltonian in the $m$th segment of the control pulse is $\hat{H}_1(t) = \hat{H}_{1,m}$ such that 

   \begin{equation}\label{eq:drives}
      \hat{H}_{1,m} =\sum_{i=1,2} \sum_{j=0}^{d-2} \sum_{\gamma=x,y}\Omega_{i,m}^{\gamma,j} \hat{\sigma}_i^{\gamma,j},
   \end{equation}

   \noindent with static Rabi frequencies $\Omega_{i,m}^\gamma$ over the duration of the segment. These frequencies will be the optimization parameters for our QuOpt protocol. For each segment, we will have eight total parameters, coming from the Rabi frequencies between neighboring transitions and $x$ and $y$ drives for each transition. Our optimizations will be over 40 total segments, leading to 320 total optimization parameters. To optimize these pulses, we use stochastic gradient descent in conjunction with PyTorch package's machine learning back propagation to calculate the gradients of the infidelity $r$ for each $\Omega_{i,m}^{\gamma,j}$. Then we use these gradients to update the Rabi frequencies with a given learning rate. This procedure lowers the infidelity and thus optimizes the control pulses. The update procedure is repeated for many iterations, leading to approximate convergence in a local minimum of the infidelity cost function landscape. Furthermore, we use the Nesterov momentum method to dynamically adjust the learning rate in the optimization \cite{nesterov1983}, which typically leads to faster convergence. An example of an optimized control pulse is shown in Fig.~\ref{fig:pulses}.

   Our QuOpt protocol described above is generally similar to the gradient ascent pulse engineering (GRAPE) algorithm  \cite{Khaneja2001,Chen2022,Petruhanov2023}, which is one of the standard techniques used to approach quantum control problems. Intuitively GRAPE starts out with a random pulse shape that is discretized into very small time steps, and then uses gradient-based optimization methods to update the static values of the pulse at the different time steps. This property suggests that the GRAPE algorithm has an advantage in finding optimized pulses, because it explores a larger space of control pulses. However, since the QuOpt protocol keeps only the frequencies that are most likely to affect the dynamics and optimizes a much smaller number of parameters, namely the Rabi frequencies in the different segments, our approach leads to significantly faster convergence compared to GRAPE. We performed GRAPE optimizations to generate the qutrit CZ gate and other gates in this paper and compare our results. These results are presented in Appendix B. 
   
\section{Empirical Analysis}
    To optimize for the qutrit CZ gate, we use the following experimental parameters. We set our coupling strength to match the coupling strength of the experimental system of Ref. \cite{Goss2022} such that $g \approx 2\pi \times 2.7$ MHz. We set the values for our qutrit frequencies as 5.4 GHz and 4.86 GHz, respectively, and the anharmonicity value as 432 MHz for both qutrits \cite{Ashhab2022}. These experimental values lead to the minimum time to implement the CZ gate or speed limit $T_{min} \approx 46$ ns, which is significantly faster than the current experimental implementation time of 580 ns in Ref. \cite{Goss2022}. See Section IV for the derivation of this speed limit. Note that the speed limit is determined primarily by the coupling strength. The qutrit frequencies, anharmonicity, and inter-qubit detuning have a small effect on the two-qutrit gate speed limit. However, these parameters, especially the anharmonicity, can play a fundamental role in determining our ability to saturate the speed limit in a real system, as we will show in Appendix C.

    For our qutrit CZ optimization, we use $M = 40$ segments with a Rabi frequency cap of $\Omega_{i,m}^{\gamma,j} \leq 40g$. This value for the Rabi frequency cap was chosen by considering what pulses would be experimentally feasible. Note that our optimized pulses typically do not reach the $40g$ cap, as seen in Fig.~\ref{fig:pulses}. We run each optimization for 5,000 iterations, and for each point, we run 50 separate random seed initializations and choose the random seed that gives the highest fidelity. Random seeds are necessary because of the sensitivity of non-convex gradient-based optimizations to the initial parameter specifications, which can lead to many optimizations ending up in sub-optimal local minima. 

    \begin{figure}[ht!]
        \centering
        \includegraphics[width=0.5\textwidth]{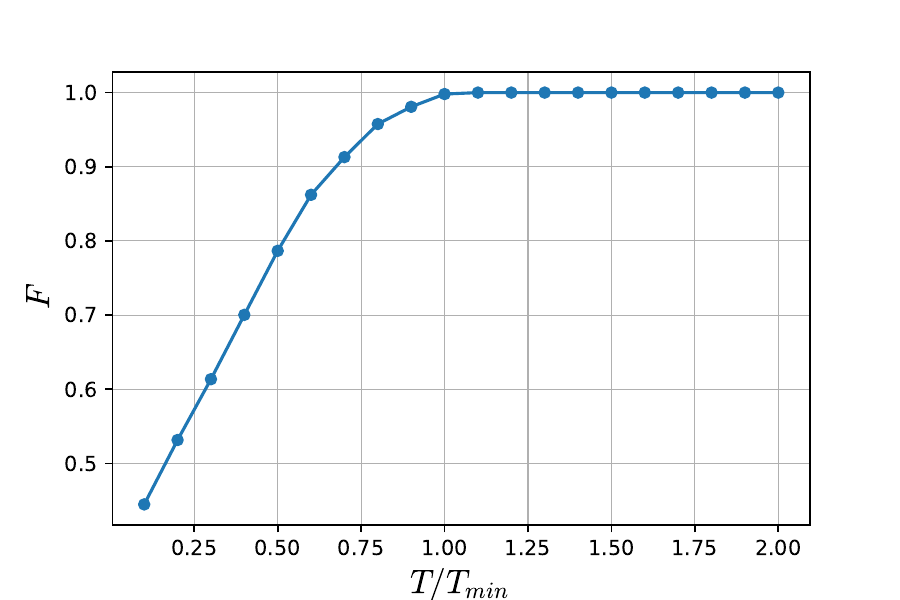}
        \caption{The fidelity $F$ of the qutrit CZ gate plotted vs. the control pulse time, expressed in terms of the speed limit of the qubit CZ gate. In other words, the time $T/T_{min} = 1$ is the minimum time needed to generate the qubit CZ gate in a two-qubit system with the same coupling strength $g$. }
        \label{fig:qutritCZ_plot}
    \end{figure}

    \begin{figure}[ht!]
        \centering
        \includegraphics[width=0.5\textwidth]{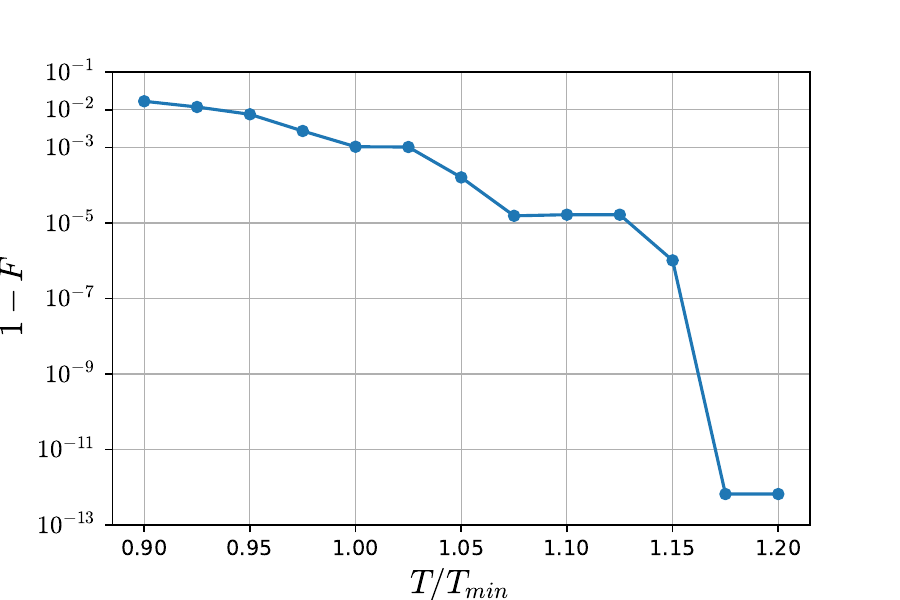}
        \caption{Infidelity $1-F$ of the qutrit CZ gate plotted versus the time to generate the gate in terms of the qubit speed limit. }
        \label{fig:logCZ}
    \end{figure}

    In Fig.~\ref{fig:qutritCZ_plot}, we plot the fidelity of the optimized gates as a function of pulse time. Here we use pulse time values in the range $[0,2*T_{min}]$, where $T_{min}$ is the minimum time needed to generate the qubit CZ gate, i.e. the qubit speed limit. This speed limit for the CZ gate is well known and studied \cite{Vidal2002,Howard2023}. See Section IV for further explanation for it's derivation. We visually observe that high-fidelity gates are obtained near the qubit CZ gate speed limit of $\sim T_{min}$. We zoom in on a small window near $T_{min},$ and we plot the results in Fig.~\ref{fig:logCZ}, where we see that high-fidelity qutrit CZ gates of $F \geq 99.9\%$ can be generated exactly at $T_{min}$. To go beyond this fidelity marker, we need to use slightly longer pulses. Specifically, to achieve a fidelity of $F\geq 99.999\%$, we need $T=1.075T_{min}$. We reach numerical precision at $T=1.175T_{min}$ such that we can presume to implement a near perfect qutrit CZ gate at this time and beyond. Thus in total, we observe that the empirical speed limit for the qutrit CZ gate is around $\sim T_{min}$. 

    The fact that the empirically derived speed limit of the qutrit CZ gate is nearly the same as that of the qubit CZ gate is intuitively surprising. One would expect that gates with a larger number of state transfers, and hence more constraints, would take longer than their counterparts in smaller Hilbert spaces. One insight on this observation is the existence of higher strength couplings when the gate protocol is expanded into higher dimensions. Note that when implementing two-qubit gates in higher-dimensional systems, it has been predicted that they can achieve a speed-up by taking advantage of the stronger couplings at higher energies and constructive interference in the coupling scheme \cite{Basyildiz2023,BasyildizGithub2023}. For example, for the iSWAP gate, one can theoretically show a 3x speed-up upon expanding the Hilbert space from a two-qubit to a two-qutrit system. Hence, implementing gates in higher-energy systems naturally leads to faster gates with highly optimized controls. Our empirical results suggest that the speed-up gained from the stronger couplings in the multi-level system balances the slow-down that might be expected from the added constraints of the qutrit gate; however, more analysis is needed to understand the factors that determine that speed limits of two-qutrit gates. 

    Our results presented above agree with those obtained using the standard GRAPE optimization method \cite{Ashhab2022,Khaneja2001} (see Appendix B). The agreement between the results of these two calculations, in spite of the significant differences in the optimization methods and cost function landscapes \cite{Chen2022,Petruhanov2023}, suggests that the observed speed limit is not a computational artifact specific to the optimization methods, but rather a more fundamental speed limit. 

    Note that the control pulse frequencies in our QuOpt calculations are set to the transition frequencies in the uncoupled two-qutrit system, where there is no hybridization between the computational basis states. The coupling between the two qutrits leads to small shifts in the energy levels and leads to hybridized energy eigenstates. Even though our QuOpt protocol does not take these corrections into account, it is successful in generating high-fidelity time-optimized gates. This robustness in the results is at least partly explained by the small coupling strength of our system compared to other energy scales in the system, leading to the shift in the transition and Rabi frequencies to be minimal. This allows us to safely ignore the effect of hybridization.

\section{Theoretical Upper Bound for Speed Limit}
    We will now derive a theoretical speed limit, expressed as a mathematical lower bound on the gate time, for the qutrit CZ gate. If both our theoretical and empirical speed limits agree, then we have found an accurate speed limit for the qutrit CZ gate and have control pulses that can saturate the bound. Note that methodologies for deriving speed limits for two-qudit gates have only recently been developed \cite{Basyildiz2023} and developing rigorous methodologies for speeds of two-qudit gates is an interesting
    open question.

    Quantum speed limits are traditionally defined in terms of orthogonal state transfer times. The time to enact a state transfer can be derived using one of two inequalities: Mandelstam-Tamm (MT) and Margolus-Levitin (ML) \cite{Frey2016,Mandelstam1991,Margolus1998}. To bound the speed for implementing a two-qubit gate, one can use the following decomposition \cite{Vidal2002}

    \begin{equation}\label{eq:cannonical}
        \hat{U} = (\hat{V_1}\otimes \hat{V_2})e^{-i\hat{H}t}(\hat{W_1} \otimes \hat{W_2}),
    \end{equation}

    \noindent where $\hat{V}_i,\hat{W}_i$ are single-site operations (for example Pauli rotations) and $e^{-i\hat{H}t}$ describes the unitary evolution generated by the Hamiltonian $\hat{H}$. The speed of the gate is defined as the time of the unitary evolution. This also indicates that the single-site operations do not factor into the speed limit. Since they are not factored into the speed limit, they can be thought of as taking zero time or instantaneously fast. Under local unitaries, we can write $\hat{H}$ in terms of a canonical formulation, $\hat{H} = h_1\hat{\sigma}_x\hat{\sigma}_x + h_2\hat{\sigma}_y\hat{\sigma}_y+h_3\hat{\sigma}_z\hat{\sigma}_z$ with ordering $h_1 \geq h_2 \geq |h_3|$ \cite{Bennett2002,Dur2001}. Using this decomposition, one can derive speed limits $T_{min}$ for common two-qubit gates such that \cite{Howard2023}

    \begin{equation}
        T_{min} = \frac{|\lambda_x| + |\lambda_y| + |\lambda_z|}{g},
    \end{equation}

    \noindent where $g > 0$ is the coupling strength between the two qubits. One relevant example of these speed limits is the one for the qubit CZ gate, where $T_{min} = \pi/4g.$ Typically speed limits for two-qubit gates are proportional to $\sim 1/g$ such that as the coupling strength between the qubits increases, the minimum time to generate the gate decreases. Speed limits of other two-qubit gates of interest can be found in \cite{Vidal2002,Howard2023,Basyildiz2024Optimal,Ashhab2012}, and the speed limits of multi-site gates and multi-qubit gates have been explored \cite{Ashhab2012,Ashhab2022,Yuan2011}. For two-qudit systems, the lower bounds on the gate times can be expressed in terms of the operator norm of the Hamiltonian $\hat{H}_I$ in the interaction picture, such that 

    \begin{equation}
        T_{min}^* \geq \frac{\pi}{2\|\hat{H_I}\|},
    \end{equation}

    \noindent where $T^*_{min}$ is the speed limit for the two-qudit gate and $\|\cdot\|$ is the operator norm. Note that this bound is built on Eq. \ref{eq:cannonical} and the state transfer bounds. To leverage this inequality, one needs to show an orthogonal state transfer for a bipartite state. One can show that the qutrit CZ gate takes the normalized $\ket{\psi} = \ket{1}\otimes(\ket{0}+\ket{1}+\ket{2})$ state to an orthogonal state such that $\bra{\psi}\hat{U}\ket{\psi} = 0$. Note that this is not the only possible orthogonal state transfer. For our capacitively coupled superconducting device Hamiltonian restricted to a qutrit space, we have $\hat{H}_I = \hat{H}_0$ and $\|\hat{H}_0\| = 3g$, such that our theoretical speed limit bound is $T_{min}^* \geq \pi/6g = 2/3T_{min}$, where again $T_{min}$ is the speed limit of the qubit CZ gate.
    
    Our bound intuitively says that the qutrit CZ gate is bounded by a time that is 1.5x faster than the qubit CZ gate; however, in our empirical results, we found a saturation of high-fidelity qutrit CZ gates around $\sim T_{min}$. This leaves a discrepancy between the empirical and theoretical speed limits. This discrepancy may be the result of a few factors. First, the optimal control results could be generating sup-optimal gates, and we cannot exclude the possibility that faster high-fidelity gates exist. However, this seems unlikely, as both GRAPE and our QuOpt results are in good overall agreement. Furthermore, the success of similar OCT approaches in generating gates near the theoretical speed limits in previous studies \cite{Basyildiz2023,Basyildiz2024Optimal,Howard2023} hints that our OCT approach is likely saturating the true speed limit. The most likely scenario is that the theoretical speed limit is loose and needs to be tightened in the future. This further drives the open question of speed limits for multi-qudit systems.
    
\section{Conclusion}
    In this paper, we investigated the speed limit of the qutrit CZ gate for a system composed of two superconducting transmons with fixed capacitive coupling. We used optimal control protocols to generate fidelity optimized qutrit CZ gates for different gate times, and identified an empirical speed limit of $T \approx T_{min}$ where $T_{min}$ is the speed limit for the qubit CZ gate. Our results are consistent across different numerical algorithms for performing optimal control, indicating that our empirical speed limit is likely a fundamental speed limit for the system under study. Furthermore, in Appendix A we show that the ququart CZ gate has the same empirical speed limit as the qubit and qutrit CZ gates. Based on these results, we conjecture that the speed limit of a general qudit CZ gate for a coupling Hamiltonian of the form in Eq.~\ref{eq:H0} is independent of $d$.

    In addition to the empirical speed limit, we derived a theoretical bound on the speed limit for the qutrit CZ gate. Our numerically optimized control protocols are slower than the theoretical bound such that a tighter theoretical bound on the speed of qutrit CZ gate may exist. An immediate future direction is to look for a tight theoretical bound saturable by our optimal protocol. One approach to develop such a bound is to generalize the method used to derive a tight speed limit for general two-qubit gates \cite{Vidal2002}. One challenge for such a generalization is that the canonical decomposition in Eq.~\ref{eq:cannonical} is only valid for two-qubit gates. Even if a similar (and likely much more complicated) decomposition can be found for a two-qutrit gate, one also has to deal with a much larger number of constraints which define a given target two-qutrit gate in order to find the corresponding speed limit. An alternative approach is to focus only the qutrit CZ gate as well as the Hamiltonian and drive defined in Eq.~\ref{eq:H0}  and Eq.~\ref{eq:control_hamiltonian} , which are experimentally relevant. In this scenario, we may be able to apply quantum speed limit bounds for some particular initial states to obtain a tight bound \cite{Frey2016,Mandelstam1991,Margolus1998}.
   
    A drawback of our approach is that we do not take into account off-resonant transitions (ORTs) such as leakage and cross-talk. Thus, our optimization is an idealized picture of a qutrit CZ gate generation but should be sufficient to obtain purely theoretical bounds on the gate time. ORTs can be included in our optimization with a more involved calculation \cite{Basyildiz2023,Basyildiz2024Optimal,BasyildizGithub2023}, and we have included optimization results using a traditional GRAPE method with ORTs in Appendix C. Note that by adding ORTs, we do see significantly longer times to generate a high-fidelity gate, nearly doubling the time. This suggests that there is likely some physical phenomenon that prevents us from saturating the theoretical speed limit. We continue this discussion in Appendix C. 

    Aside from theoretical analysis, an experimental implementation of fidelity- and speed-optimized qutrit CZ gates is a natural application of our work. While qutrit CZ gates have been demonstrated on superconducting hardware, speed-optimized versions of these gates are yet to be implemented \cite{Goss2022,Luo2023}.  Similar speed optimizations has been successfully demonstrated for two-qubit gates \cite{Howard2023}. These speed- and fidelity-optimized gates can have a significant impact on the practical application of current and future areas of quantum computation, as they can enable achieving higher practical circuit depths.

\section*{Acknowledgments}
    We thank John Preskill for enlightening discussions on qutrit speed limits. We thank the Research Computing HPC cluster of the University of Colorado Boulder for the computational resources. BB acknowledges funding provided by the Institute for Quantum Information and Matter, an NSF Physics Frontiers Center (NSF Grant PHY-2317110). SA acknowledges support from Japan's Ministry of Education, Culture, Sports, Science and Technology (MEXT) Quantum Leap Flagship Program Grant Number JPMXS0120319794.

\section*{Appendix A: Speed limits of ququart CZ gate}
    As another step in extending our analysis to multi-level CZ gates, we look at the empirical speed limit of the ququart CZ gate. Note that the qudit CZ gate is defined as \cite{Wang2020,Goss2022}

    \begin{equation}
        CZ\ket{k}\ket{l} = e^{i2\pi k l /d}\ket{k}\ket{l}
    \end{equation}

    \noindent where $d$ is the number of levels in each qudit ($d=2$ for qubits). For the ququart gate, we have $d = 4$ such that $CZ|_{d=4}\ket{k}\ket{l} = e^{i\pi k l/2}\ket{k}\ket{l} $ for $k,l\in\{0,1,2,3\}.$ We use our QuOpt optimization to generate high-fidelity and speed-optimized ququart CZ gates. Compared to our protocols for the qutrit CZ gate, we will add control drives for the $\ket{2}\leftrightarrow\ket{3}$ transition such that our drive term $H_{1,m}$ of the Hamiltonian $H = H_0 + H_{1,m}$ is same as Eq. \ref{eq:control_hamiltonian} but with $d=4.$ 
    
\begin{figure}[ht!]
        \centering
        \includegraphics[width=0.5\textwidth]{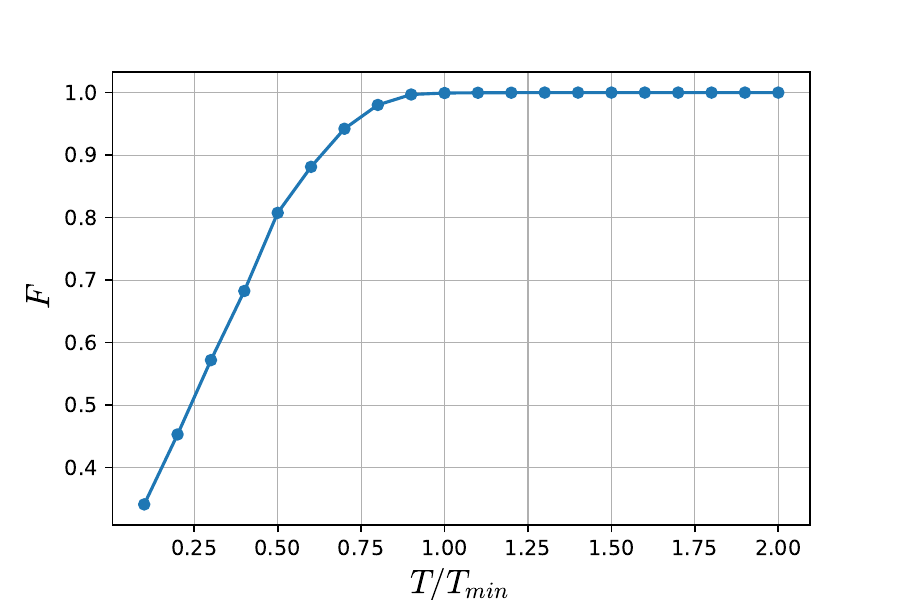}
        \caption{The fidelity $F$ of the ququart CZ gate plotted vs. the control pulse time, expressed in terms of the speed limit of the qubit CZ gate. In other words, the time $T/T_{min} = 1$ is the minimum time needed to generate the qubit CZ gate in a two-qubit system with the same coupling strength $g$. }
        \label{fig:CZ4}
    \end{figure}

    Here $\hat{H}_0$ is the capactive coupling Hamiltonian in Eq.~\ref{eq:H0} and static drives applied in a segmented fashion for $M = 40$ total segments. Fixing the time $T$, we run our optimization to minimize infidelity $r =1-F$ with the fidelity metric described in Eq.~\ref{eq: fidelity} with $d = 16$ and $\hat{U}_j = \hat{V}_k \otimes \hat{W}_l$, where $\hat{V}_k,\hat{W}_l$ are the generators for $SU(4).$ Now running our optimization over the same timescale as Fig. \ref{fig:qutritCZ_plot}, we obtain the results shown Fig. \ref{fig:CZ4}.
    
\begin{figure}[ht!]
        \centering
        \includegraphics[width=0.5\textwidth]{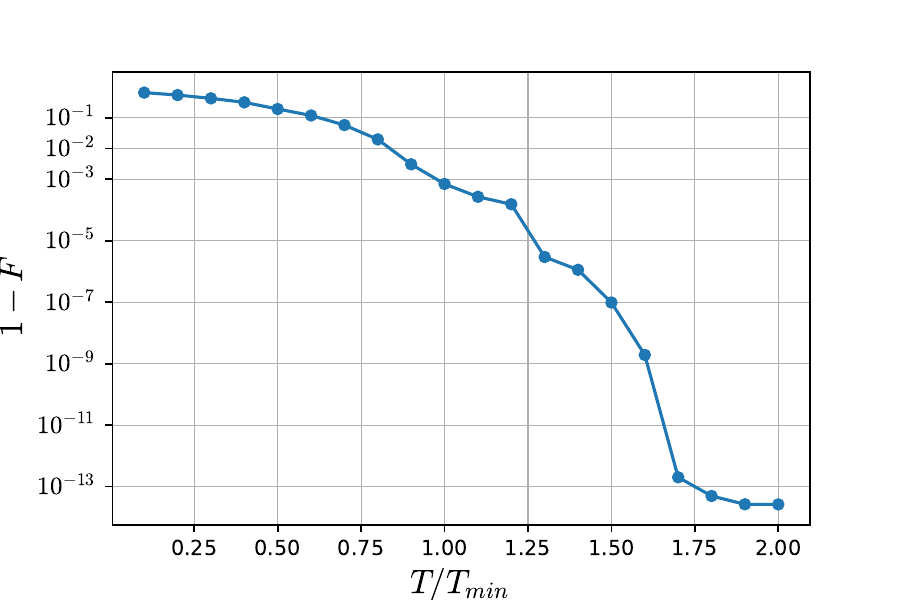}
        \caption{Linear-log plot of the infidelity $1-F$ of the ququart CZ gate versus the control pulse time. This figure uses the same data plotted in Fig.~\ref{fig:CZ4}.} 
        \label{fig:logCZ4}
    \end{figure}

     For each value of $T$, we use 50 random seed initializations for the control parameters, and each optimization is run for 5,000 iterations. In Fig. \ref{fig:CZ4}, we see that we can generate the ququart CZ gate at $\sim T_{min}$, similar to the qutrit CZ gate. Plotting the infidelity on a logarithmic scale vs. time, we obtain the results in Fig. \ref{fig:logCZ4}. 

    We see that in Fig. \ref{fig:logCZ4}, a high-fidelity $(F \geq 99.9\%)$ ququart CZ gate can be generated at $T_{min}$. This result is again intuitively surprising. As in the case of the qutrit CZ gate, it seems that there is a balance between the increase in $g$ from higher energy couplings and the increase in the number of constraints as $d$ gets larger.

\section*{Appendix B: Alternative optimization algorithm}
\begin{figure}[ht!]
        \centering
        \includegraphics[width=0.5\textwidth]{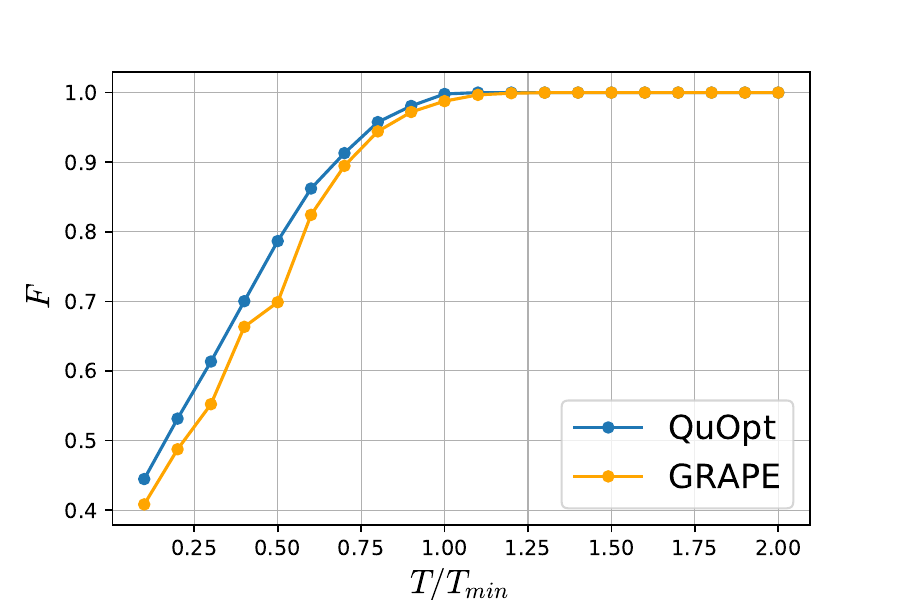}
        \caption{The fidelity $F$ of the qutrit CZ gate plotted vs. the control pulse time, for the QuOpt protocol vs. GRAPE for the qutrit CZ gate.}
        \label{fig: Comp}
    \end{figure}

    In this appendix, we compare our QuOpt optimization with those obtained using the gradient ascent pulse engineering (GRAPE) algorithm \cite{Khaneja2001,Chen2022,Petruhanov2023}. Intuitively, GRAPE starts out with a random pulse that is discretized into a piece-wise constant function with very small time steps, and then uses gradient-based methods to optimize the static values of the pulse in the different time steps. Our GRAPE calculations divide the total time $T$ into $10^4$ time steps and run for $10^4$ optimization iterations. We use 10 random initializations for each value of $T$ and take the maximum fidelity. The results are plotted in Fig.~\ref{fig: Comp}. 

\begin{figure}[ht!]
        \centering
        \includegraphics[width=0.5\textwidth]{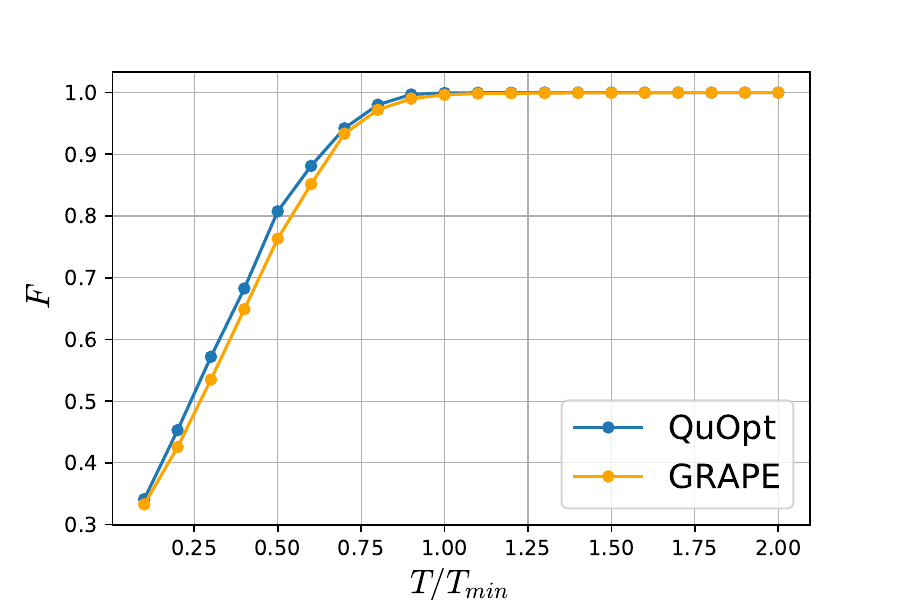}
        \caption{The fidelity $F$ of the qutrit CZ gate plotted vs. the control pulse time, for the QuOpt protocol vs. GRAPE for the ququart CZ gate.}
        \label{fig: Comp_CZ4}
    \end{figure}

    We can see in Fig. \ref{fig: Comp} that the GRAPE results have only slightly lower high fidelity values than those obtained with our native code-base. The speed limit is essentially the same in both cases: $T \approx T_{min}$. Repeating this process for the ququart CZ gate, we obtain the results in Fig. \ref{fig: Comp_CZ4}

    Again, the GRAPE results largely match those obtained with our native code-base. The fact that two different optimization schemes give essentially the same results increases our confidence in the accuracy of the numerical results. 

\section*{Appendix C: Off-resonant transitions}
    So far, since we are mainly interested in the relation between the speed limit and the qudit-qudit coupling strength, we have assumed full control on the individual qudits. In particular, we have assumed that the different transitions in each qudit can be addressed individually. In practice, there are a few typical error sources, and these include leakage to higher-energy states and cross-talk between qudits. We refer to these error sources as off-resonant transitions (ORTs). In this appendix, we include these error sources in the theoretical model and analyze their effect on the two-qudit-gate speed limit.
    
    We first note that we will show only the results obtained using the GRAPE algorithm. This is due to step size errors in the time evolution dynamics in the QuOpt code-base, that produce non-realistic results when ORTs were introduced. We plan to include results using an improved version of that approach in a future version of the manuscript. Note that this error does not affect any of the results reported in this paper, as these errors derive from the discretizations of time steps in the ORT protocol of the QuOpt optimizations. When ORTs are introduced, they create time-dependent phase differences, even in the interaction frame. Thus, one has to use numerical ODE methods to account for the time dependence, which results in step-size errors in the numerical ODE solution. Our GRAPE results are not affected by this issue.
    
    The main and almost-inevitable ORT effect is the one in which the driving field that is intended to target a certain transition in a qudit also acts as an off-resonant driving field for other transitions of the same qudit. From an experimental point of view, we can say that there is only one drive signal applied to each qudit, and all frequency components are included in this one signal. This description also suggests that we do not have separate terms for the real and imaginary parts of the signal; any distinction between the real and imaginary parts would be implemented by applying a phase shift to the applied signal. Considering typical superconducting qubit designs, ORT effects can be modeled by approximating each qudit as a truncated weakly anharmonic oscillator and writing the control Hamiltonian as

    \begin{equation}\label{eq:CZ4_Drives}
        \hat{H}_1(t) = \sum_{j=0,1} \Omega_j(t) \left( \hat{a}_j + \hat{a}_j^{\dagger} \right),
    \end{equation}

    \noindent where $\hat{a}_j$ and $\hat{a}_j^{\dagger}$ are, respectively, the annihilation and creation operators of qudit $j$. These operators are defined by the standard harmonic oscillator formula $\bra{k}a\ket{l}=\sqrt{l}\delta_{k,l-1}$. If we compare Eq.~\ref{eq:CZ4_Drives} with Eq.~\ref{eq:control_hamiltonian}, we can see that there is only one control field per qudit in Eq.~\ref{eq:CZ4_Drives}, compared to the $2(d-1)$ control fields per qudit in Eq.~\ref{eq:control_hamiltonian}.
    
\begin{figure}[ht!]
        \centering
        \includegraphics[width=0.5\textwidth]{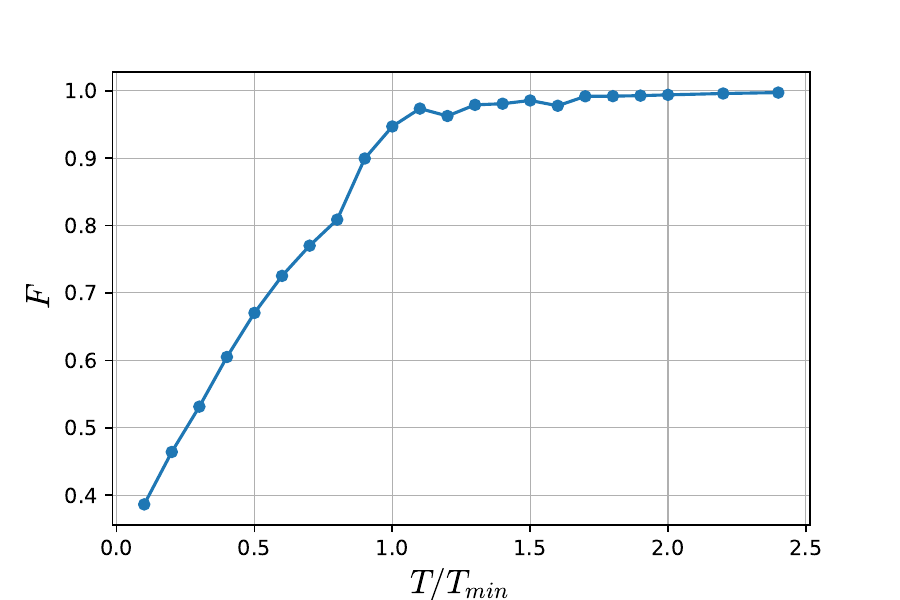}
        \caption{The fidelity $F$ of the qutrit CZ gate plotted vs. the control pulse time, using GRAPE optimization and including ORTs.}
        \label{fig: GRAPE ORT CZ3}
    \end{figure}

    Now, optimizing for the qutrit CZ gate using GRAPE, we obtain the results plotted in Fig.~\ref{fig: GRAPE ORT CZ3}.

    In Fig.~\ref{fig: GRAPE ORT CZ3} we achieve a high-fidelity gate at $T > 2.5T_{min}$, and we see that these error sources significantly impact our ability to generate the target gate, nearly tripling our time to generate a high-fidelity gate. This slowdown may be partly due to having a more complex cost function landscape, leading to our optimizer having difficulty converging to a quality local minimum. However, the slowdown almost certainly has a more fundamental origin, stemming from limited controllability of the system when ORTs are taken into account. In other words, the theoretical speed limits in the main text obtained without any error sources are likely unachievable for a realistic experimental system. 

\begin{figure}[ht!]
        \centering
        \includegraphics[width=0.5\textwidth]{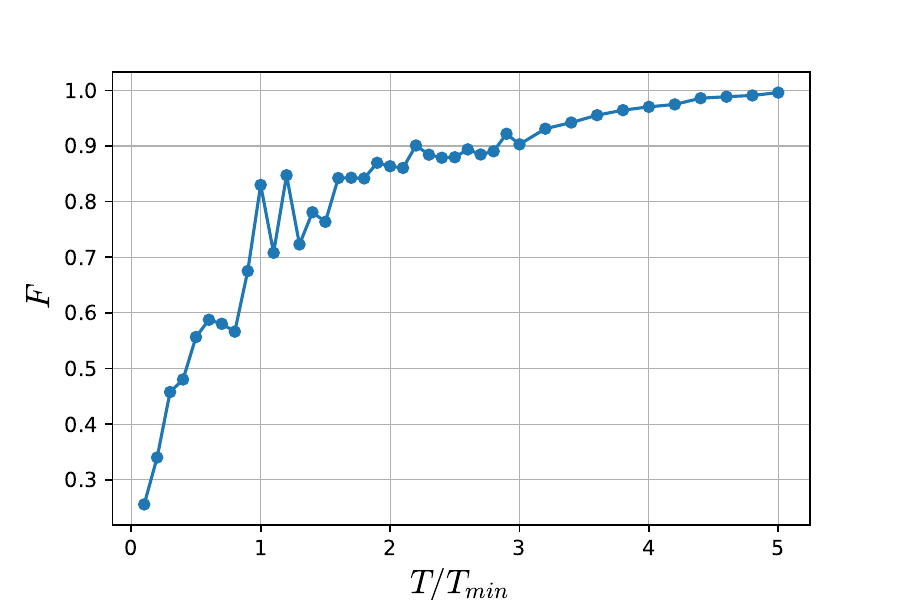}
        \caption{The fidelity $F$ of the ququart CZ gate plotted vs. the control pulse time, using GRAPE optimization and including ORTs.}
        \label{fig: GRAPE ORT CZ4}
    \end{figure}

    Repeating our optimization for the ququart CZ gate while including the ORT effects, we obtain Fig.~\ref{fig: GRAPE ORT CZ4}.

    Here we see a more significant slowdown. A high-fidelity gate is achieved only at $T > 5T_{min}$, near an order of magnitude slowdown compared to our idealized control model where a high-fidelity gate was achieved at $T \approx T_{min}$. As in the qutrit case, this slowdown may indicate that our optimizer is less efficient, as the higher level of complexity introduced by the ququart system may lead to difficulty in parameter convergence. However, we expect that the slowdown is a physical effect caused by the rather restricted control capabilities in the realistic model.

\bibliographystyle{apsrev4-2}
\bibliography{refs}

\end{document}